\documentclass[sn-mathphys]{sn-jnl}

\usepackage{amsmath,amstext}
\usepackage[T1]{fontenc}
\usepackage{graphicx}
\usepackage{xcolor}
\definecolor{ForestGreen}{HTML}{2e8b21}
\definecolor{NatureRed}{HTML}{f54e42}
\definecolor{NatureBlue}{HTML}{4254f5}
\definecolor{NatureGreen}{HTML}{138f57}
\definecolor{NatureMagenta}{HTML}{6b1345}
\definecolor{NaturePurple}{HTML}{c227b0}
\definecolor{NatureCyan}{HTML}{2b807b}
\definecolor{NatureOrange}{HTML}{f5a142}
\usepackage{appendix}
\usepackage{bookmark} 
\usepackage{array,multirow}
\usepackage{amssymb}
\usepackage{hyperref}
\usepackage{subcaption}
\usepackage{textgreek}
\usepackage{xspace}
\usepackage{booktabs}
\usepackage{multirow}
\usepackage[symbol]{footmisc}
\usepackage{lineno}

\jyear{2023}

\theoremstyle{thmstyleone}%
\theoremstyle{thmstyletwo}%

\theoremstyle{thmstylethree}%
\newcommand{\ssim}{\ensuremath{\sim\!}\xspace}
\newcommand{\lthan}{\ensuremath{\!<\!}\xspace}

\newcommand{\re}{\ensuremath{R_\mathrm{e}}\xspace}

\newcommand{\Mstar}{\ensuremath{M_\star}\xspace}
\newcommand{\tquench}{\ensuremath{\mathrm{t_{quench}}}\xspace}
\newcommand{\tform}{\ensuremath{\mathrm{t_{form}}\xspace}}

\newcommand{\MSun}{\ensuremath{{\rm M}_\odot}\xspace}
\newcommand{\Met}{\ensuremath{\mathrm{log_{10}(Z/Z_{\odot})}}\xspace}
\newcommand{\fluxcgs}{\ensuremath{\mathrm{erg\,s^{-1}\,cm^{-2}}}\xspace}
\let\oldAA\AA
\renewcommand{\AA}{\oldAA\xspace}
\let\oldtextsigma\textsigma
\renewcommand{\textsigma}{\oldtextsigma\xspace}

\newcommand{\targetidfull}{JADES-GS+53.15508-27.80178\xspace}
\newcommand{\targetid}{JADES-GS-z7-01-QU\xspace}

\newcommand{\fesc}{\ensuremath{f_{\rm esc}}\xspace}

\newcommand{\LymanAlpha}{\text{Ly\textalpha}\xspace}

\newcommand{\Hbeta}{\text{H\textbeta}\xspace}
\newcommand{\Hdelta}{\text{H\textdelta}\xspace}
\newcommand{\OIII}{\text{[O\,{\sc iii}]\textlambda5008}\xspace}

\newcommand{\jwst}{\textit{JWST}\xspace}

\newcommand{\ppxf}{{\sc ppxf}\xspace}
\newcommand{\bagpipes}{{\sc bagpipes}\xspace}
\newcommand{\beagle}{{\sc beagle}\xspace}
\newcommand{\prospector}{{\sc prospector}\xspace}
\newcommand{\forcepho}{{\sc forcepho}\xspace}

\newcolumntype{L}[1]{>{\raggedright\let\newline\\\arraybackslash\hspace{0pt}}m{#1}}
\newcolumntype{C}[1]{>{\centering\let\newline\\\arraybackslash\hspace{0pt}}m{#1}}
\newcolumntype{R}[1]{>{\raggedleft\let\newline\\\arraybackslash\hspace{0pt}}m{#1}}

\begin{document}

\title[A recently quenched galaxy 700 Myr after the Big Bang]{A recently quenched galaxy 700 million years after the Big Bang}

\author*[1,2]{\fnm{Tobias} \sur{J.~Looser}}\email{tjl54@cam.ac.uk}
\author[1,2]{\fnm{Francesco} \sur{D'Eugenio}}
\author[1,2,3]{\fnm{Roberto} \sur{Maiolino}}
\author[1,2]{\fnm{Joris} \sur{Witstok}}
\author[1,2]{\fnm{Lester} \sur{Sandles}}
\author[4]{\fnm{Emma} \sur{Curtis-Lake}}
\author[5]{\fnm{Jacopo} \sur{Chevallard}}
\author[1,2]{\fnm{Sandro} \sur{Tacchella}}
\author[6]{\fnm{Benjamin} \sur{D.~Johnson}}
\author[1,2]{\fnm{William} \sur{M.~Baker}}
\author[]{\fnm{Katherine} \sur{A.~Suess$^{\textit{\normalfont{7,8}}}$}} 

\author[9]{\fnm{Stefano} \sur{Carniani}}
\author[10]{\fnm{Pierre} \sur{Ferruit}}
\author[11]{\fnm{Santiago} \sur{Arribas$^{\textit{\normalfont{11}}}$}} 
\author[12,13]{\fnm{Nina} \sur{Bonaventura}}
\author[5]{\fnm{Andrew} \sur{J.~Bunker}}
\author[5]{\fnm{Alex} \sur{J. Cameron}}
\author[14]{\fnm{Stephane} \sur{Charlot}}
\author[1,2,15]{\fnm{Mirko} \sur{Curti}}
\author[16]{\fnm{Anna} \sur{de Graaff}}
\author[17]{\fnm{Michael} \sur{V.~Maseda}}
\author[18]{\fnm{Tim} \sur{Rawle}}
\author[16]{\fnm{Hans-Walter} \sur{Rix}}
\author[11]{\fnm{Bruno} \sur{Rodríguez Del Pino$^{\textit{\normalfont{11}}}$}} 
\author[19]{\fnm{Renske} \sur{Smit}}
\author[1,2]{\fnm{Hannah} \sur{\"Ubler}}
\author[20]{\fnm{Chris} \sur{Willott}}

\author[21]{\fnm{Stacey} \sur{Alberts}}
\author[21]{\fnm{Eiichi} \sur{Egami}}
\author[6]{\fnm{Daniel} \sur{J.\ Eisenstein}}
\author[22]{\fnm{Ryan} \sur{Endsley$^{\textit{\normalfont{22}}}$}} 
\author[23]{\fnm{Ryan} \sur{Hausen}}
\author[21]{\fnm{Marcia} \sur{Rieke}}
\author[7]{\fnm{Brant} \sur{Robertson}} 
\author[21]{\fnm{Irene} \sur{Shivaei}}
\author[24]{\fnm{Christina} \sur{C.~Williams}}

\author[25]{\fnm{Kristan} \sur{Boyett}}
\author[21]{\fnm{Zuyi} \sur{Chen}}
\author[21]{\fnm{Zhiyuan} \sur{Ji}}
\author[5]{\fnm{Gareth} \sur{C. Jones}}
\author[26]{\fnm{Nimisha} \sur{Kumari}}
\author[27]{\fnm{Erica} \sur{Nelson}}
\author[11]{\fnm{Michele} \sur{Perna$^{\textit{\normalfont{11}}}$}} 
\author[3,5]{\fnm{Aayush} \sur{Saxena}}
\author[1,2]{\fnm{Jan} \sur{Scholtz}}

\affil[1]{Kavli Institute for Cosmology, University of Cambridge, Madingley Road, Cambridge, CB3 OHA, UK}

\affil[2]{Cavendish Laboratory - Astrophysics Group, University of Cambridge, 19 JJ Thomson Avenue, Cambridge, CB3 OHE, UK}

\affil[3]{Department of Physics and Astronomy, University College London, Gower Street, London WC1E 6BT, UK}

\affil[4]{Centre for Astrophysics Research, Department of Physics, Astronomy and Mathematics, University of Hertfordshire, Hatfield AL10 9AB, UK}

\affil[5]{Department of Physics, University of Oxford, Denys Wilkinson Building, Keble Road, Oxford OX1 3RH, UK}

\affil[6]{Center for Astrophysics $\|$ Harvard \& Smithsonian, 60 Garden St., Cambridge, MA 02138, USA}

\affil[7]{Department of Astronomy and Astrophysics University of California, Santa Cruz, 1156 High Street, Santa Cruz CA 96054, USA}

\affil[8]{Kavli Institute for Particle Astrophysics and Cosmology and Department of Physics, Stanford University, Stanford, CA 94305, USA}

\affil[9]{Scuola Normale Superiore, Piazza dei Cavalieri 7, I-56126 Pisa, Italy}

\affil[10]{European Space Agency, European Space Astronomy Centre, Camino Bajo del Castillo s/n, 28692 Villafranca del Castillo, Madrid, Spain}

\affil[11]{Centro de Astrobiolog\'ia (CAB), CSIC–INTA, Cra. de Ajalvir Km.~4, 28850- Torrej\'on de Ardoz, Madrid, Spain}

\affil[12]{Cosmic Dawn Center (DAWN), Copenhagen, Denmark} 

\affil[13]{Niels Bohr Institute, University of Copenhagen, Jagtvej 128, DK-2200, Copenhagen, Denmark}

\affil[14]{Sorbonne Universit\'e, CNRS, UMR 7095, Institut d'Astrophysique de Paris, 98 bis bd Arago, 75014 Paris, France}

\affil[15]{European Southern Observatory, Karl-Schwarzschild-Strasse 2, D-85748 Garching bei Muenchen, Germany}

\affil[16]{Max-Planck-Institut f\"ur Astronomie, K\"onigstuhl 17, D-69117, Heidelberg, Germany}

\affil[17]{Department of Astronomy, University of Wisconsin-Madison, 475 N. Charter St., Madison, WI 53706, USA}

\affil[18]{European Space Agency (ESA), ESA Office, STScI, Baltimore, MD 21218, USA}

\affil[19]{Astrophysics Research Institute, Liverpool John Moores University, 146 Brownlow Hill, Liverpool L3 5RF, UK}

\affil[20]{NRC Herzberg, 5071 West Saanich Rd, Victoria, BC V9E 2E7, Canada}

\affil[21]{Steward Observatory University of Arizona 933 N. Cherry Avenue ,Tucson, AZ 85721, USA}

\affil[22]{Department of Astronomy, University of Texas, Austin, TX 78712, USA}

\affil[23]{Department of Physics and Astronomy, The Johns Hopkins University,  3400 N. Charles St., Baltimore, MD 21218, USA}

\affil[24]{NSF's National Optical-Infrared Astronomy Research Laboratory, 950 North Cherry Avenue, Tucson, AZ 85719, USA}

\affil[25]{School of Physics, University of Melbourne, Parkville 3010, VIC, Australia}

\affil[26]{AURA for European Space Agency, Space Telescope Science Institute, 3700 San Martin Drive, Baltimore, MD 21218, USA}

\affil[27]{Department for Astrophysical and Planetary Science, University of Colorado, Boulder, CO 80309, USA}


\abstract{
\textbf{Local and low-redshift ($z$<$3$) galaxies are known to broadly follow a bimodal distribution: actively star forming galaxies with relatively stable star-formation rates, and passive systems. These two populations are connected by galaxies in relatively slow transition.
In contrast, theory predicts that star formation was stochastic at early cosmic times and in low-mass systems \citep{Endsley+2022,Tacchella+2023,Mason+2023,Whitler+2023}: these galaxies transitioned rapidly between starburst episodes and phases of suppressed star formation, potentially even causing temporary quiescence -- so-called mini-quenching events \citep{Endsley2023,Dome+2023}. However, the regime of star-formation burstiness is observationally highly unconstrained.
Directly observing mini-quenched galaxies in the primordial Universe is therefore of utmost importance to constrain models of galaxy formation and transformation \citep{Merlin2019MNRAS.490.3309M, Lovell2022arXiv221107540L}.
Early quenched galaxies have been identified out to redshift $z \lthan 5$ \cite[e.g.][]{Glazebrook2017Natur.544...71G, 
Carnall2022arXiv220800986C, 
Valentino2020ApJ...889...93V,
Nanayakkara2022arXiv221211638N}, and these are all found to be massive ($\Mstar>10^{10}~\MSun$) and relatively old.
Here we report a (mini-)quenched galaxy at z$=$7.3, when the Universe was only 700~Myr old. 
The \jwst/NIRSpec spectrum 
is very blue ($U$-$V$$=$0.16$\pm0.03$~mag), but exhibits a Balmer break and no nebular emission lines.
The galaxy experienced a short starburst followed by rapid quenching; its stellar mass
(4-6$\times 10^8~M_\odot$) falls in a range that is sensitive to various feedback mechanisms, which can result in perhaps only temporary quenching. 
}}



\maketitle

The galaxy was first described 
as a Lyman break galaxy \cite[e.g.][]{Oesch2010ApJ...709L..16O}; and was recently observed as part of our \jwst Advanced Deep Extragalactic Survey (JADES; galaxy ID: \targetidfull, hereafter simply \targetid) 
through deep (28~h) NIRSpec-MSA observations with the prism. The galaxy was pre-selected with 
the photometric Lyman dropout technique and a blue rest-frame UV colour.

\begin{figure}
  \centering
  \includegraphics[width=1\textwidth]{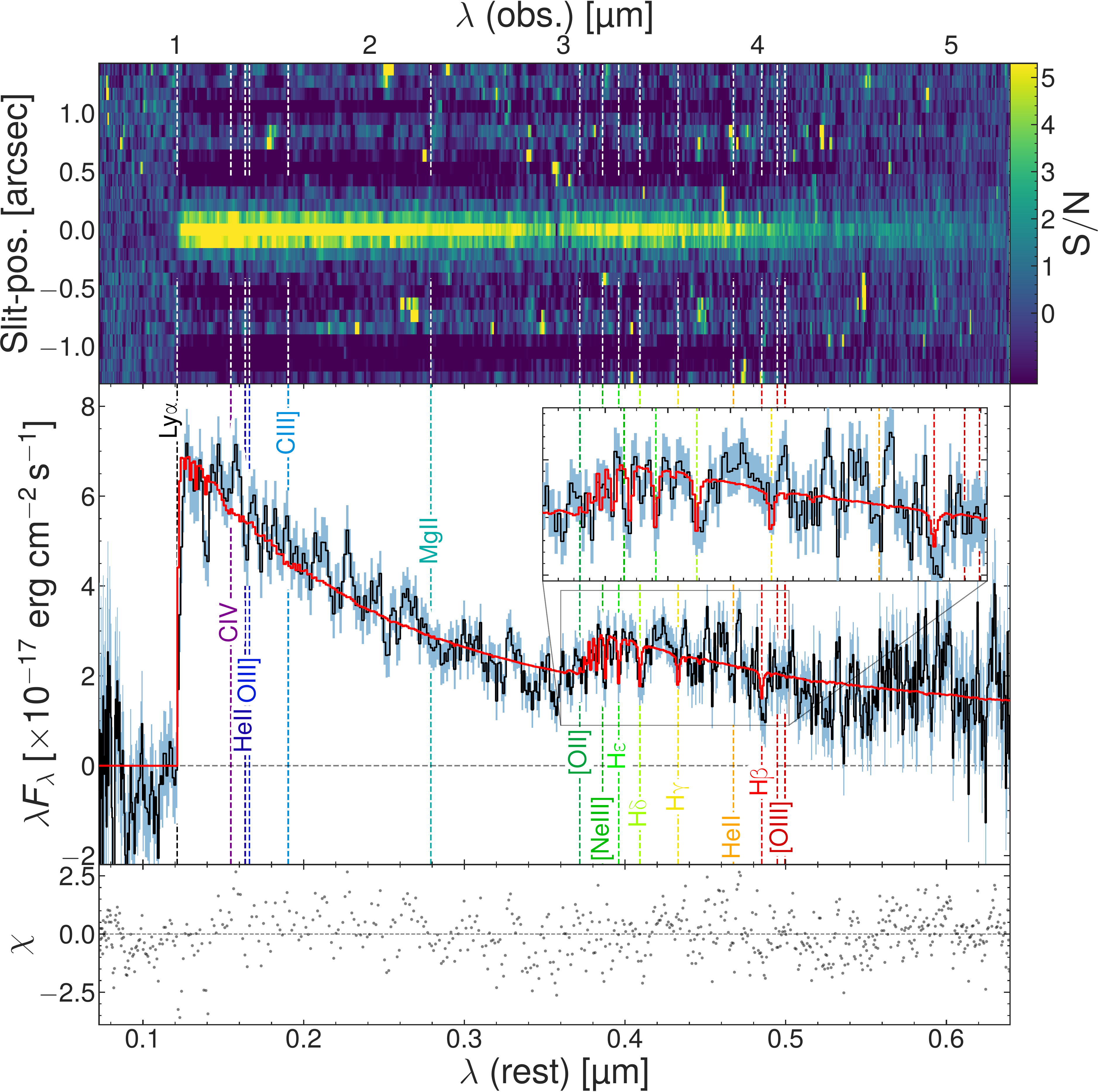}
  \caption{\textbf{NIRSpec R100/prism spectrum of \targetid.
  The absence of emission lines, together with the Balmer break, reveals that this is a -- temporarily or permanently -- (mini-)quenched, post-starburst galaxy.} The clearly detected \LymanAlpha drop and the Balmer break unambiguously give a redshift of z=7.3.
The vertical dashed lines indicate the rest-frame wavelengths of the strongest nebular emission lines.
The red line indicates the \ppxf spectral fit. The upper panel shows the signal-to-noise ratio in the 2D prism spectrum. The bottom panel shows the ratio between the residuals of the fit and the noise. For reference, the flux in the F200W NIRCam filter is $3.33 \pm 0.08 \times 10^{-17}~{\rm erg} \ {\rm cm}^{-2} \ {\rm s}^{-1}$, fully consistent with the spectrum.}
  \label{pPXF_fit}
\end{figure}

The spectrum of \targetid is shown in Fig.\ref{pPXF_fit}.
The redshift $z$=$7.29\pm 0.01$ is unambiguously determined (using the \beagle code, see Methods) from the combined observed wavelengths of the characteristic \LymanAlpha drop and Balmer break.

The 3-\textsigma upper limit on the \Hbeta emission-line flux, $F(\Hbeta) \lthan 6.1\times 10^{-20}~{\rm erg} \ {\rm cm}^{-2} \ {\rm s}^{-1}$,  implies an upper limit on the star-formation rate (SFR) of~ $\lthan 0.65$~\MSun~yr$^{-1}$ over the last 3-10~Myr (even accounting for dust attenuation, see Methods).
Even stronger constraints come from the \OIII line: we find $F(\OIII)\lthan 6.5\times10^{-20}$~\fluxcgs, which, combined with a conservative assumption on the \OIII/\Hbeta ratios in high-$z$ galaxies \citep{Cameron2023arXiv230204298C,sanders2023arXiv230106696S}, implies a 3-\textsigma limit on the SFR five times lower than the \Hbeta-derived value. 
The absence of emission lines is
independently confirmed by the medium-resolution spectrum (see Methods).

We measure a UV slope $\beta=-2.09\pm0.09$, typical for galaxies at $6\lthan z \lthan 10$ \citep{Bhatawdekar2021ApJ...909..144B,Bunker2023arXiv230207256B}, indicating strong star-formation activity during the last 100 Myr before observation. 
In the rest-frame visible, we detect a clear Balmer break and \Hdelta absorption with
equivalent width ${\rm EW_{\Hdelta_A}} = 4.8\pm1.0$~\AA. This value,
-- combined with the absence of emission lines -- means that \targetid meets the most common 
spectroscopic definition of a post-starburst galaxy \citep{Goto2007MNRAS.381..187G%
, wu2018ApJ...868...37W}
, i.e., a galaxy that has only recently
stopped forming stars.

Previous high-redshift works have identified Balmer-break galaxies 
in the Epoch of Reionization
\cite[e.g.][]{Hashimoto2018Natur.557..392H,Roberts-Borsani2020MNRAS.497.3440R, Laporte2021MNRAS.505.3336L}
, indicating the existence of evolved stellar populations and even proposing quiescent phases in these objects \cite[e.g.][]{Hashimoto2018Natur.557..392H,Roberts-Borsani2020MNRAS.497.3440R}. However, without spectroscopy, one cannot rule out the presence of emission lines with low equivalent
width or that strong emission lines masquerade as Balmer breaks. Additionally, due to the lack of atmospheric transmission at wavelengths longer than 2.5 microns, it is impossible to investigate Balmer breaks at $z>5$ from the ground. Therefore, before \jwst, it was impossible to confirm the absence of ongoing star formation.

Crucially, based on colours alone, this (mini-)quenched galaxy would have been identified as `star forming' by the colour selection criteria \citep{Williams+2009}, even if including the extension to fast-quenched galaxies \citep{Belli+2019}. Indeed, its rest-frame
$U$-$V$ colour of 0.16$\pm0.03$~mag places it outside the quiescent region
of the $UVJ$ diagram, regardless of $V$-$J$ colour
\citep{park+2022}, as it is the case for other quiescent galaxies at high redshift \citep{Merlin2019MNRAS.490.3309M}.
However, thanks to \jwst/NIRSpec, we can place stringent upper limits on the nebular emission line fluxes.

Are there potential alternatives to the quenched interpretation? A very high escape 
fraction of ionising Lyman-continuum (LyC) photons with \fesc$>$ 0.9 could strongly suppress nebular emission \citep{Zackrisson2017ApJ...836...78Z}
. However, if \fesc is high, this would be because nearly all of the ISM was ejected or consumed by star formation \citep{Rosdahl2022MNRAS.515.2386R}; yet, if the ISM is absent, there is no fuel for star formation and the galaxy must be quenched. This makes the galaxy highly interesting in the context of reionization, as a remnant leaker \citep{Katz2023MNRAS.518..270K}.
The question is whether the object is (still) a remnant leaker at the epoch of observation. In other words, whether there are still very young stellar populations (a few Myr old) which would still be producing ionizing photons associated with O-type stars, and which would largely escape the galaxy, as \fesc$\approx1$. This scenario is disfavoured by the normal UV slope $\beta$ \citep{Topping2022ApJ...941..153T}, the  
Balmer break and by the strong \Hdelta absorption.

Statistically, a very recently (< 10 Myr) star-forming solution with high \fesc is also
disfavoured by our further analysis. Indeed, by leveraging on the flexibility of the 
software \beagle to model the observed spectrum, we find 
that a high-\fesc, recently SF solution -- though possible -- is 
strongly disfavoured compared to the quenched (> 3-10 Myr)solution (see 
Methods). Additionally, as we will discuss below, both the \ppxf and \prospector codes, which can optionally decouple the continuum from the nebular lines (which are degenerate with $\rm f_{esc}$) do not favour a solution with very recent star formation.
The second alternative that we cannot completely rule out is the presence of completely obscured star formation,
as advocated for some post-starburst galaxies in the local Universe \citep{Baron2022arXiv220411881B}.
However, we note that high dust masses and high dust extinction in such low mass systems, at such high redshift, have never been observed \citep{Sandles+2023a}.

\medskip

To estimate the physical properties of the galaxy including stellar mass \Mstar, SFR, SFH, dust attenuation and stellar metallicity, we apply joint spectro-photometric modelling of its spectral energy distribution (SED).
To marginalise over model assumptions and implementation, we use four different SED-fitting codes (\ppxf 
;
\bagpipes; 
\prospector
; and \beagle
, see Methods).
Fig.~\ref{pPXF_fit} shows, as an example, the best-fit \ppxf model in red, overlaid on the spectrum.

\begin{figure}[ht]
\centering
  \includegraphics[width=1\textwidth]{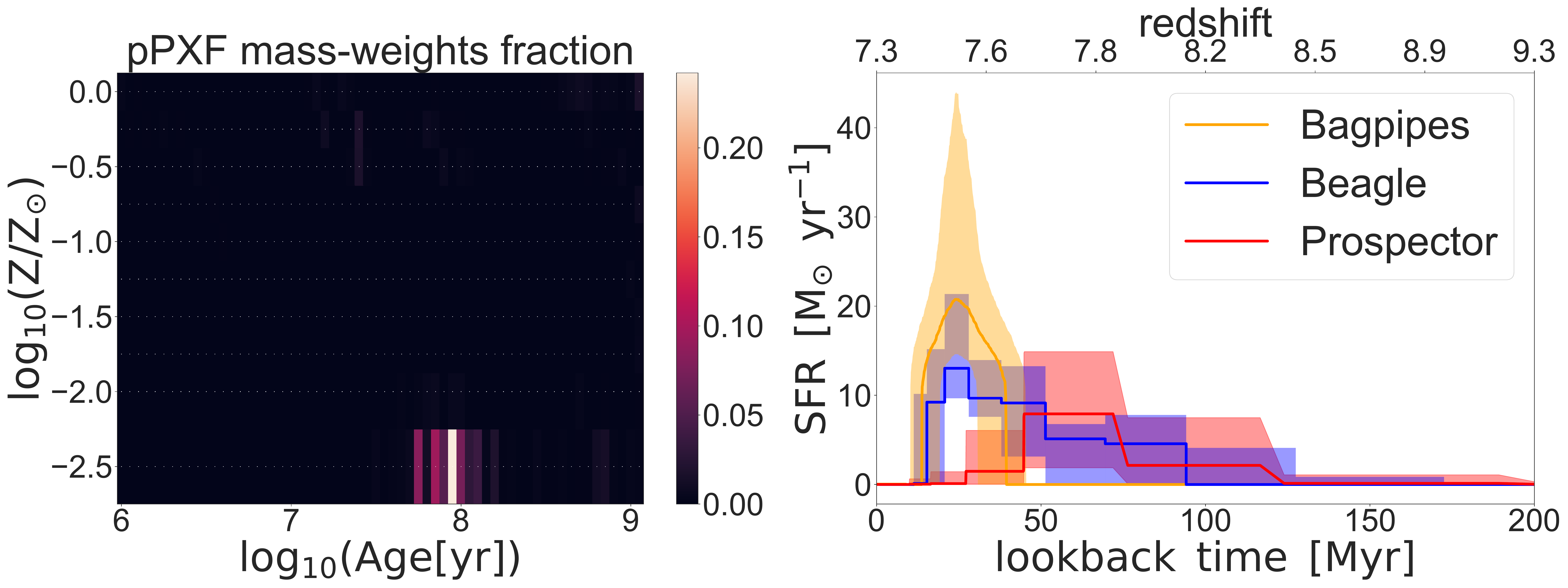}
        \caption{\textbf{The galaxy's star-formation history as inferred by four different full spectral fitting codes, using different -- effective -- priors.}
        All four codes confirm that the galaxy is quenched at the epoch of observation and reconstruct comparable SFHs. Left: The stellar age-metallicity grid inferred by \ppxf. The code reconstructs dominant metal-poor populations forming from $\sim$150 to $\sim$50~Myr prior to observation. The colour-bar represents the fractional mass distribution over the SSP grid. Right: 
         The star formation histories inferred by \bagpipes, \beagle and \prospector. \bagpipes infers that the galaxy formed $\sim$40~Myr before the epoch of observation and quenched $\sim$20 Myr before observation. \beagle infers that the galaxy quenched $\sim$20~Myr before observation after a star-burst lasting $\sim$70~Myr. \prospector infers that the galaxy quenched $\sim$40~Myr before observation after a star-burst lasting $\sim$80~Myr.}
        \label{fig:assembly_histories}
\end{figure}

\begin{table}[h]
    \centering
\begin{tabular}{lrrrrr}
\hline
\hline
Key inferred properties & \ppxf & \bagpipes & \beagle & \prospector
\\
\hline
\hline
log$_{10}$(\Mstar/$M_{\odot}$) & -& $8.5 \pm0.1$ & $8.8^{+0.1}_{-0.2}$ & $8.7^{+0.1}_{-0.1}$ \\
log$_{10}$(SFR [$M_{\odot}$/yr])& - & $<\text{--}1.0$ & \text{--}$2.5^{+1.0}_{-1.0}$ &  \text{--}$2.6^{+1.5}_{-2.7}$ \\
\Met & $<\text{--}2.0$ & $-0.7 \pm 0.1 $& \text{--}$1.9^{+0.4}_{-0.2}$ & \text{--}$1.7^{+0.2}_{-0.2}$\\
\tquench [Myr] & $\sim50$ & $18^{+5}_{-5}$ & $16^{+7}_{-4}$ &$38^{+9}_{-10}$\\
\tform [Myr] & $\sim150$& $37^{+8}_{-5}$ & $93^{+69}_{-47}$ & $116_{-45}^{+ 85}$\\
$\mathrm{A_V}$ [mag] & $0.4 \pm 0.1$ & $0.32^{+0.26}_{-0.23}$ & $0.51^{+0.03}_{-0.04}$ & $0.1^{+0.1}_{-0.0}$\\
\hline
\hline

\

\end{tabular}
\caption{\textbf{Key physical quantities inferred by the four full spectral fitting codes \ppxf, \bagpipes, \beagle and \prospector}. From top to bottom: Stellar mass (\Mstar), Star-formation rate (SFR),  metallicity (Z), quenching lookback time (\tquench), formation lookback time (\tform), and effective dust attenuation optical depth ($A_V$). 
}
\label{tab:Basic_outputs}
\end{table}
\

The methods agree on a low stellar mass of \Mstar=~4--6$\times10^8$~\MSun
(Table~\ref{tab:Basic_outputs}), i.e. this is an object in the dwarf-galaxy regime -- essentially the same mass as the nearby, actively star-forming Small Magellanic Cloud, but at z=7.3 and quenched.

Fig.~\ref{fig:assembly_histories} shows the SFH of the galaxy, as inferred by the four codes. All models agree that \targetid is quenched and give similar stellar population parameters. The oldest significant population of stars is 40--150~Myr old, corresponding
to a formation redshift $z$=7.6--8.8, 
while the youngest stars have ages 20--50~Myr, corresponding to a quenching redshift of $z$=7.4--7.6. These numbers imply that \targetid
formed in a burst of star formation lasting only 20--100~Myr -- consistent with the 
formation timescales of star-forming galaxies
at similar redshifts \citep{Tacchella2022arXiv220803281T}.

The SFR at the time of observation inferred by \bagpipes, \beagle and \prospector are extremely low, between $10^{-2.6}$ and $10^{-1.4}~\MSun~\rm{yr}^{-1}$, yielding specific SFRs ranging between $10^{-2.3}$ ${\rm Gyr}^{-1}$ and $0.1~ \rm Gyr^{-1}$; these values are between 2 and 3 orders of magnitude below the Main Sequence of star-forming galaxies at this redshift \cite[e.g.][]{Steinhardt2014ApJ...791L..25S,Pearson2018A&A...615A.146P,Sandles2022MNRAS.515.2951S, Santini2017ApJ...847...76S,Laporte2022arXiv221205072L}
, and below the widely used threshold ${\rm sSFR_{10}} < 0.2/{\rm t_{H}} = 0.29 \ {\rm Gyr}^{-1}$, on 10 Myr timescales, hence confirming that the galaxy is quenched at the epoch of observation. Crucially, the four codes agree that galaxy has been strongly star-forming between 10 and 100 Myr prior to the epoch of observation.

Three of the four codes infer a tentative low average stellar metallicity of the galaxy of \Met $\approx -2$ (where $Z_{\odot}$ is the solar metallicity), while \bagpipes infers \Met $\approx -0.7$. \ppxf indicates the presence of a weak enriched population representing only 5 percent of the total stellar mass of the galaxy, which formed last before quenching. However, we note that stellar metallicity measurements are uncertain with the low-resolution prism spectroscopy. 
\medskip

Which physical mechanism(s) quenched the galaxy? 

The inferred mass of this galaxy rules out that it has been quenched by the UV background \citep{Efstathiou1992MNRAS.256P..43E}; indeed, numerical simulations predict that this quenching mechanism works only for very low-mass galaxies with $\Mstar \approx10^5-10^7~\MSun$ (maximally $<10^8~\MSun$) \citep[e.g.][]{Katz2020MNRAS.494.2200K
}.
In the local Universe, galaxies in the mass range of our target are quenched primarily by environment \citep{peng+2010, bluck+2020b}. 
And it has been postulated that some satellite galaxies may experience environment-driven quenching already during the Epoch of Reionization \citep{Gelli2021ApJ...913L..25G}.
However, we do not find any massive galaxies nearby (see Methods), disfavoring environmental effects as quenching mechanism for this target.

Given the short inferred duration of the SFH and the rapidity of the transition to quiescence, it seems more reasonable
to speculate that \targetid may have experienced a powerful outflow, driven either by star-formation feedback (radiation-pressure, supernovae might act too slowly) or by accretion on a primeval supermassive black hole, which rapidly ejected most of the
star-forming gas \citep{Koudmani2019MNRAS.484.2047K}. This scenario is supported by the tentative low average stellar metallicity inferred by three of the codes. Indeed, ejective feedback mechanisms might have rapidly removed gas from the galaxy and quenched it, before the ISM could be significantly enriched with new metals. A slower quenching process (such as the starvation scenarios) would have probably resulted in a longer
transition between star forming and quenched, and into higher-metallicity stellar populations, formed out of 
recycled gas produced by stellar evolution and returned to the ISM via supernovae \citep{Peng2015Natur.521..192P,Trussler2020MNRAS.491.5406T
}. 

These outflow events, 
either SF or AGN driven, might have mini-quenched star formation only temporarily \citep{Ceverino2018MNRAS.480.4842C}, 
until new or re-accreted material replenishes the supply of gas available for star formation and rejuvenates the 
galaxy.
The latter picture may be qualitatively in agreement with a wide range of cosmological simulations predicting that a population of galaxies in the early Universe goes through periodic bursts of star formation, interspersed with periods of suppressed star formation \cite[e.g.][]{Kimm2015MNRAS.451.2900K,Ma2018MNRAS.478.1694M,Ceverino2018MNRAS.480.4842C}. Although the expected SFHs are very `bursty', these recent simulations struggle to achieve the complete quenching observed by us for galaxies with mass similar to our system. 

More generally, interpreting these observations with existing simulations is complicated because, according to current theories \citep{Ma2018MNRAS.478.1694M}, this object occupies the transition region between bursty and stable SFHs. Moreover, it is important to note that these models do not include AGN feedback, which recent observations have
shown to be important in local galaxies of this mass range \cite[e.g.][]{Penny2018MNRAS.476..979P}
. These difficulties mean that \targetid
provides the community with the opportunity to shed light on this pivotal mass range.

We conclude by emphasizing that the discovery and spectroscopic analysis of a (mini-)quenched galaxy at redshift z$=$7.3 by our JADES
collaboration ushers the era in which we can constrain theoretical feedback models using direct observations of the primordial Universe.
However, this is just the starting point for the \jwst~mission: upcoming and future observations will start the
transition from the `discovery' phase to the statistical characterization of the properties of the first (mini-)quenched galaxies.

\clearpage
\renewcommand{\figurename}{Extended Data Fig.}
\renewcommand{\tablename}{Extended Data Table}
\setcounter{figure}{0}
\setcounter{table}{0}

\newpage

\section{Methods}
\subsection{\jwst/NIRSpec spectra}
The NIRSpec \citep{Jakobsen2022A&A...661A..80J} prism/R100 and gratings/R1000 spectra of \targetid presented in this work were obtained as part of our JADES GTO programme (PI: N. Lützgendorf, ID:1210) observations in the Great Observatories Origins Deep Survey South (GOODS-S) field between October 21--25, 2022. 
The R100 observations were obtained using the disperser/filter configuration
{\sc prism/clear}, which covers the wavelength range between 0.6~µm and 5.3~µm and provides spectra with a wavelength dependent spectral resolution
of $R \ssim$30--330. The R100 spectrum of \targetid is presented in Fig.~\ref{pPXF_fit}.

\begin{figure}
  \centering
  \includegraphics[width=1\textwidth]{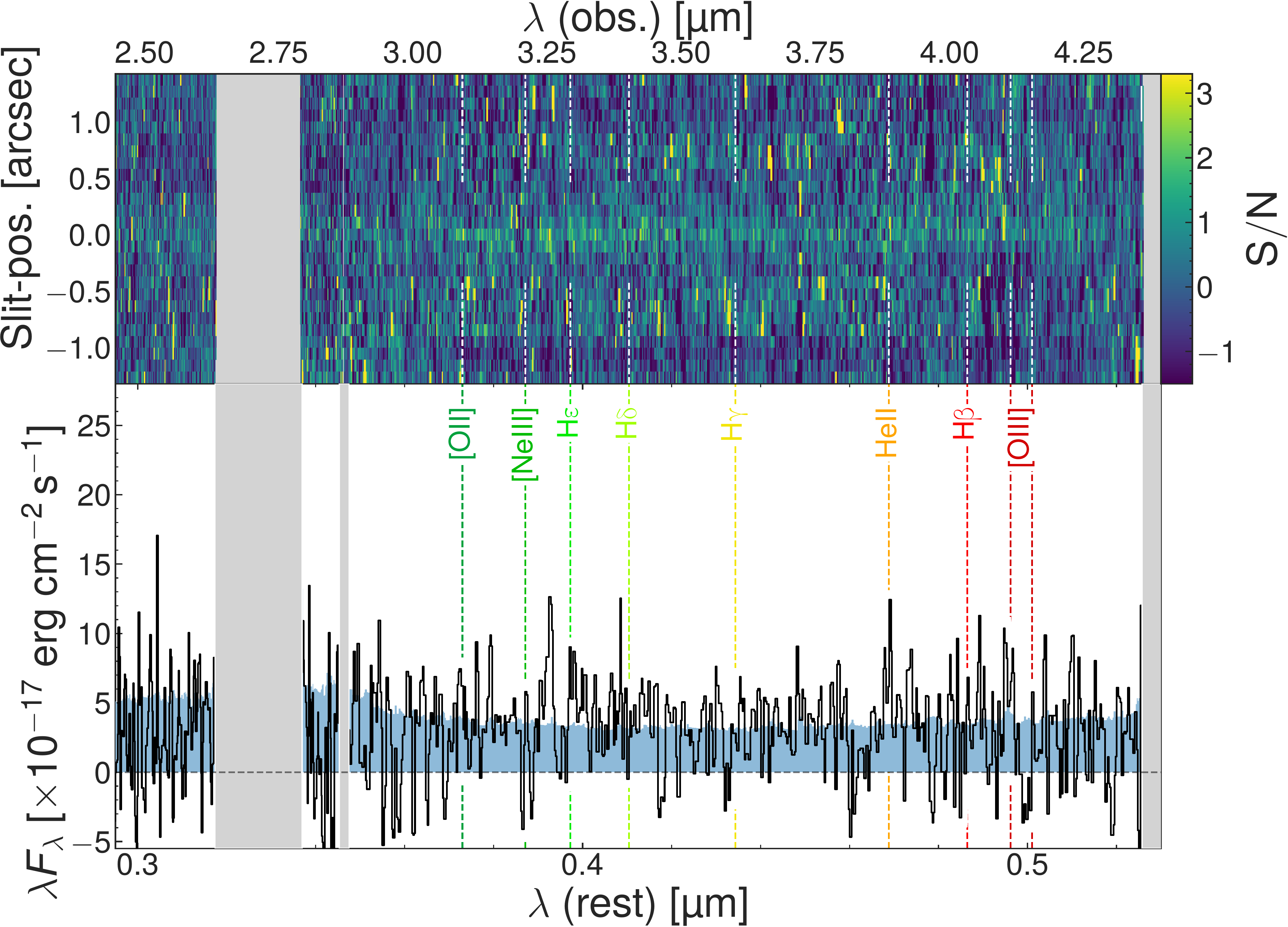}
  \caption{{\bf NIRSpec R1000/grating spectrum of the (mini-)quenched galaxy \targetid at z = 7.3}. The spectrum confirms the absence of emission lines. The blue shaded region shows the 1D noise level. The upper panel shows the signal-to-noise ratio in the 2D grating spectrum. The spectrum is median-smoothed, for visualisation. 
  }
  \label{pPXF_R1000}
\end{figure}

The medium resolution R1000 observations, with a spectral resolution of $R \ssim$500--1340 used the disperser/filter configurations {\sc G140M/F070LP}, {\sc G235M/F170LP} and {\sc G395M/F290LP}, which were exposed for 14~h, 7~h
and 7~h. A zoom-in on the R1000 spectrum (into the region with spectral lines best tracing star-formation activity) is shown in Extended Data Fig.~\ref{pPXF_R1000}. Finally, high-resolution R2700 observations used {\sc G395H/F290LP} and
were exposed for 7~h (like the R1000 spectrum, the R2700 spectrum of \targetid contains no detections hence is not shown).

The programme observed a total of 253 galaxies over three dither pointings, with \targetid has been observed in each of the three pointings. Each dither pointing had a different microshutter array (MSA) configuration to place the spectra at different positions on the detector - to decrease the impact of detector gaps, mitigate detector artefacts and improve the signal-to-noise ratio for high-priority targets, while increasing the density of observed targets. Within each individual dither pointing the telescope executed a three nod pattern (by slightly re-orienting the telescope by the length of one microshutter, keeping the same MSA configuration).  In each of the three nodding pointings, three microshutters were opened for each target, with the targets in the central shutter.
Each three-point nodding was executed within 8403 seconds. The nodding pattern has been repeated four times in the prism/clear configuration, two times in the G140M/F070LP combination, once in the G235M/F170LP combination and once in the G395M/F290LP combination. This resulted in a total exposure time for \targetid of 28 hours in R100, 14 hours in  G140M, and 7 hours in each of G235M, G395M and G395H.

The flux-calibrated spectra were extracted using a customized pipeline developed by the NIRSpec GTO Team, which builds on the publicly available ESA NIRSpec Science Operations Team (SOT) pipeline \cite{Ferruit2022}. A detailed description of the custom pipeline will be presented in a forthcoming technical paper (Carniani et al., in preparation), and more information can be found in \citep{Bunker2023_HST_DEEP_overview}. We summarize here the main steps and the differences to the publicly available pipeline. For each exposure we extract the count rate for each pixel, removing cosmic rays, and flagging saturation. The 2D spectrum is background subtracted based on the two other exposures in the three-nod pattern. 
The individual 2D spectra are flat
fielded and illumination-corrected taking into account the wavelength-dependent throughput. The wavelength and flux calibration was then applied,
with each pixel of the 2D spectrum having an associated wavelength and position along the shutter. We applied a wavelength-dependant path-loss correction to account
for flux falling outside the micro-shutter, taking into account the considerable PSF variation of NIRSpec, treating the target as a point source. For the prism, we used an irregular spectral wavelength grid, taking into account the resolution (R) as a function of wavelength. The 1D spectra for the three nod positions from each of the three pointings are combined by a weighted average into a single 1D spectrum. Outliers are rejected with a sigma clipping algorithm.
The presented 1D
spectra come from a combination of the 1D individual spectra, and are not an extraction from the presented combined 2D spectra.

\subsection{\jwst/NIRCam image and morphology}

\begin{figure}
  \centering
  \includegraphics[width=1\textwidth]{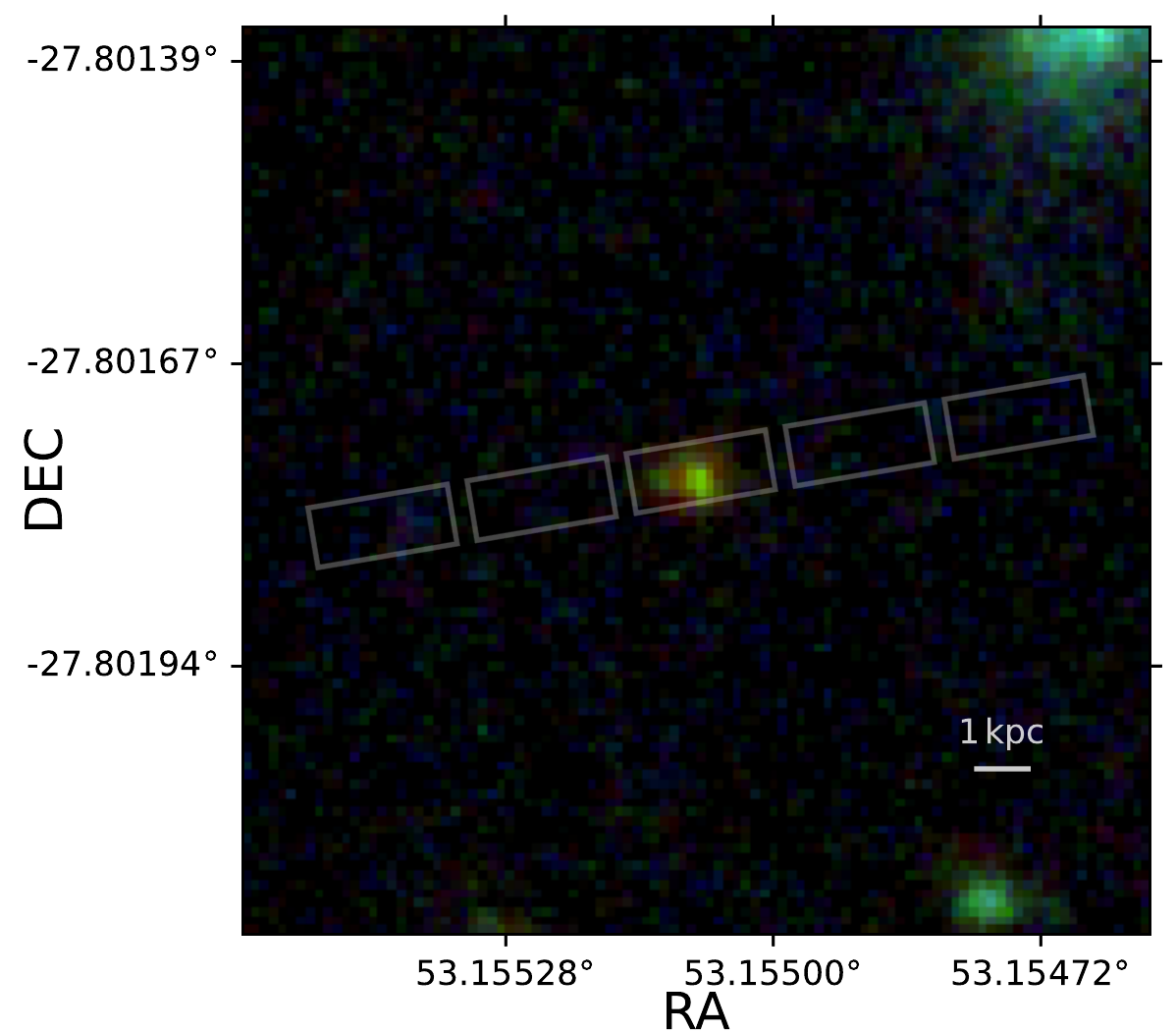}
  \caption{{\bf \jwst/NIRCam image covering \targetid and its nearby projected environment.} The NIRCam F444W-F200W-F090W rgb colour image is created from cutouts of the mosaics at wavelengths \textlambda $\approx\!0.8$--5~\textmu ms. The five NIRSpec microshutter positions used for this target are overlaid in white.}\label{fig:NIRCam}
\end{figure}

A \jwst/NIRCam F444W-F200W-F090W rgb (red-green-blue) colour image of \targetid from our JADES programme (PI: Daniel J. Eisenstein, ID:1180), created from cutouts of the mosaics in each filter, at wavelengths $\lambda \approx 0.8 - 5~\mu$m, is shown in Extended Data Fig.~\ref{fig:NIRCam}.

For the spectro-photometric modelling of \targetid we used the photometry from the JADES and JEMS \citep{Williams2023arXiv230109780W} NIRCam \citep{Rieke2005SPIE.5904....1R, Rieke2022arXiv221212069R} surveys. In particular, the modelling included deep infrared NIRCam observations with the following filters: F090W, F115W, F150W, F182M, F200W, F210M, F277W, F335M, F356W, F410M, F430M, F444W, F460M and F480M. The JADES photometry reduction pipeline made use of the JWST Calibration Pipeline (JWSTCP,
v1.9.2) with the CRDS pmap context 1039. The raw images were transformed into count-rate images, making use of the JWSTCP stage 1, where detector-level corrections and `snowballs' were accounted for. The count-rate images were then flat fielded and flux calibrated with a customised methodology, using JWSTCP stage 2. Finally, the mosaics were created using the stage 3 of the pipeline. For further details on the JADES photometry data reduction pipeline we refer to Robertson et al. 2022 \citep{Robertson2022arXiv221204480R} and Tacchella et al. 2023 \citep{Tacchella2023arXiv230207234T}.

To obtain the morphological parameters of \targetid, we fit the NIRCam photometry with \forcepho (Johnson et~al., in~prep.). \forcepho models galaxies and substructures (e.g. clumps or blended companions) as multiple Sérsic profiles convolved with the instrument PSFs as mixtures of Gaussians by forward modelling the light distribution in all individual exposures and filters and sampling the joint posterior probability distribution of all parameters via MCMC. For more details on the multi-component modelling procedure, see Tacchella et al. 2023 \citep{Tacchella2023arXiv230207234T}. \targetid appears as a compact, discy galaxy (half-light
radius \re= $36\pm1$~mas $\widehat{=}$ 0.2~kpc $\widehat{=}$ 0.04 arcsec, S\'ersic index $n=0.95\pm0.03$; Extended Data Fig.~\ref{fig:NIRCam}). The
images also show a distinct, fainter source 0.13~arcsec to the East.
This secondary source could not be deblended in the spectroscopy, but we 
obtained deblended photometry using \forcepho.
The contribution of the secondary source to the total flux ranges from a maximum of 27~percent
(in the F115W band) to 17~percent (F444W band), therefore its SED is 
significantly bluer than that of the main source.
Its photometric redshift z=7.50$\pm0.13$ (1-\textsigma) is consistent with the spectroscopic redshift of the main
source. At a redshift of $z$=$7.3$, this secondary source would lie within 
0.7~kpc (or 3~\re) from the centre of \targetid; its interpretation as a
clump or satellite is unclear.
To attempt removing its contribution from the spectrum of the main source,
we extracted a spectrum from the central three pixels (0.3 arcsec) from the NIRSpec 2-d 
spectrum; using this spectrum does not change the interpretation of our 
results, i.e., \targetid is still quenched.

As discussed in the main text, quenching by environment is ruled out for \targetid, as no other galaxy resides nearby. This can be verified with JADES NIRCam imaging on our publicly available website; and more specifically the interactive tool FitsMap: \url{https://jades.idies.jhu.edu/public/?ra=53.1554497&dec=-27.8018917&zoom=9} at the coordinates RA=53.1551 and DEC= -- 27.8018.

\subsection{Full spectral fitting}

\subsubsection{\ppxf}\label{sec.ppxf}
The red model fit of the stellar continuum in Fig.~\ref{pPXF_fit} was performed with the $\chi^2$-minimization Penalized PiXel-Fitting code 
\ppxf \citep{cappellari2017, cappellari2022}, using a library of single stellar-population (SSP) templates spectra obtained combining the synthetic C3K model atmospheres \citep{Conroy2019} with MIST isochrones \citep{Choi2016} and solar abundances. The SSP spectra span a full 2D logarithmic grid of 60 ages and 10 metallicities from age$_{SSP}$ = $10^{6.0}$~yr to $10^{9.2}$~yr (generously older than the age of the Universe at z$=$7.3) and $\Met_{SSP}$ = -2.5 to 0.5. Due to the low resolution of the R100 spectrum, we fix the stellar velocity dispersion to its virial estimate $\sigma_{\rm *} \approx \sigma_{\rm vir} \equiv \sqrt{G M_*/(5R_{\rm e})} = \mathrm{50~km/s}$. To account for dust reddening, the fitted SSP are multiplicatively coupled to the Calzetti et al. (2000) \cite{Calzetti2000ApJ...533..682C} dust attenuation curve. To infer the stellar population weight-grid shown in the left panel of Fig.~\ref{fig:assembly_histories}, following \cite{Looser2023b}, we first convolve the SSP templates to match the
wavelength-dependant spectral resolution of the prism spectrum. Then, to avoid numerical problems, both the spectrum and the templates are re-normalized by the
median flux per spectral pixel. Then we run an initial fit with \ppxf, and we $\sigma$-clip outliers in the spectrum. Finally, we perform a residual-based bootstrapping of the initial \ppxf bestfit, without regularization \citep{cappellari2017, cappellari2022}, over 1000 iterations. The inferred bootstrapped SSP-grids are averaged to recover the non-parametric SFH, consistent with the intrinsic noise of the spectrum, presented in the left panel of Fig.~\ref{fig:assembly_histories}. 

We infer a dust attenuation of the stars in this galaxy of $A_V=0.4 \pm 0.1$. It should be noted that the presence of dust in the \ppxf fit is mainly driven by the UV slope. The complex physics of the \LymanAlpha drop is not included in the SSP templates. Masking this part of the spectrum returns a nearly dust-free fit with older and metal-richer stellar populations, which would make \targetid even more
quenched. As stated in the main text, we infer an extremely low average stellar metallicity of \Met$\approx-2$ with \ppxf. It should be noted that the dominant reconstructed stellar populations lie at \Met$ \approx -2.5$, at the boundary of the available grid of synthetic spectra. This suggests that model SSP spectra of even lower metallicity might be needed in the future to accurately model the stellar populations in galaxies at high redshift. However, we note that the metallicity measurements are uncertain, due to the low resolution of the prism. We infer that the oldest significant population of stars (i.e. indicating the start of the SF) in the galaxy is 150~Myr old, while the youngest is 50~Myr, resulting in an extremely short duration of the star formation of just 100~Myr between the formation of the galaxy and its quenching.

\subsubsection{\bagpipes}\label{sec.bagpipes}
\begin{figure}
  \centering
  \includegraphics[width=1\textwidth]{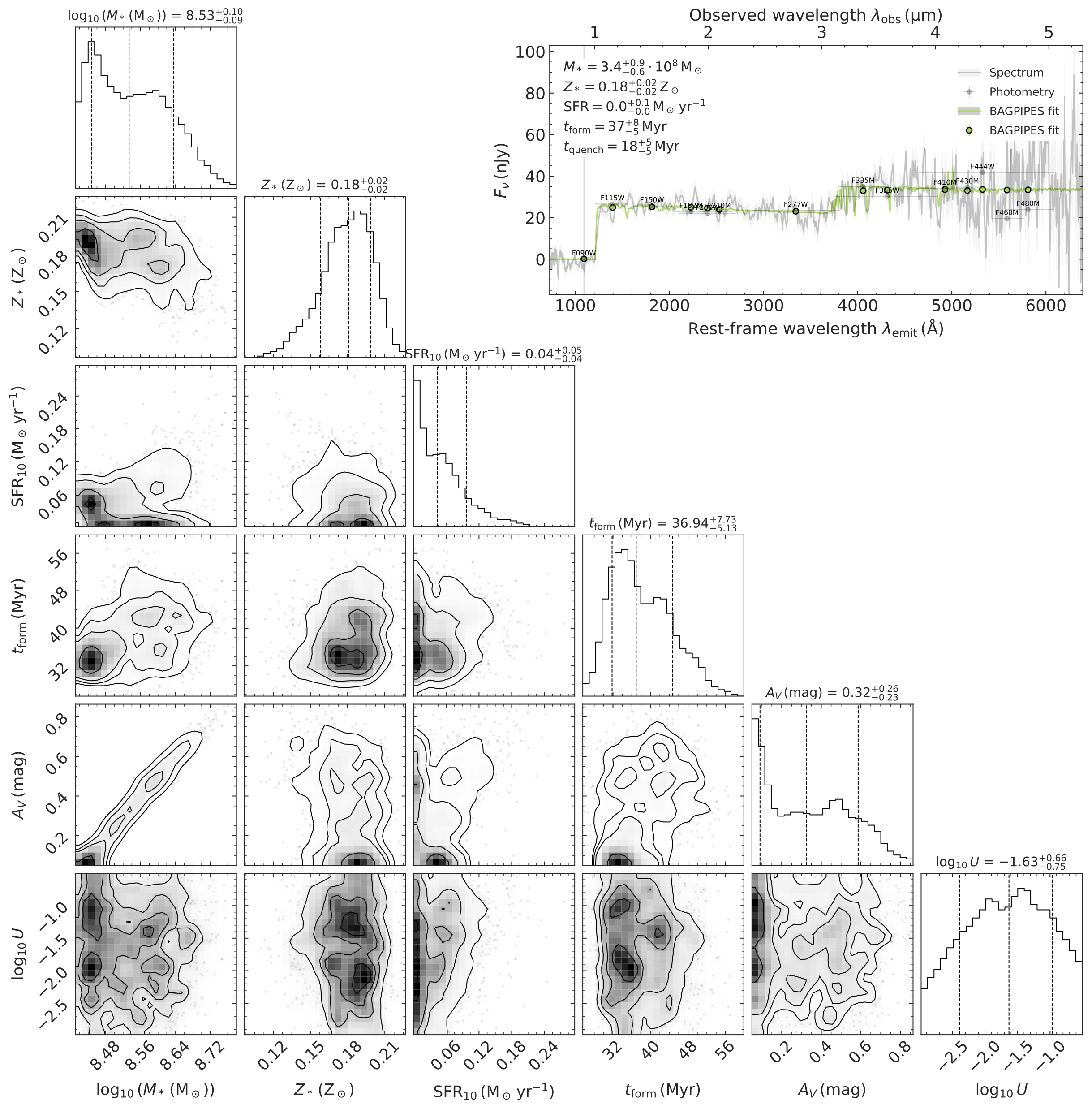}
  \caption{{\bf Summary of key outputs by \bagpipes.} Bottom left: corner plot. Top right: spectro-photometric \bagpipes fit of the \targetid R100/prism spectrum.}
  \label{Bagpipes_corner_plot}
\end{figure}

We used the Bayesian Analysis of Galaxies for Physical Inference and Parameter EStimation (\textsc{bagpipes}) code \cite{Carnall2018MNRAS.480.4379C} to simultaneously fit the NIRSpec PRISM measurements and NIRCam photometry. Following Witstok et al. 2023 \cite{2023arXiv230205468W}, we employed the updated \textsc{BC03} stellar population
models \citep{bc2003,Chevallard&Charlot2016} combined with the stellar MILES library \citep{Sanchez+2006} and the updated stellar evolution tracks \citep{Bressan+2012,Marigo+2013}. For the presented \bagpipes fit, we assumed two bins of constant SFH, one fixed bin over the last 10~Myr and one variable bin spanning a range beyond 10~Myr (minimum age ranging between 10~Myr and 0.5~Gyr, maximum age between 11~Myr and the age of the Universe). We varied the total stellar mass formed between $0$ and $10^{15} \, \mathrm{M_\odot}$, and the stellar metallicity of the variable SFH bin between $0.01 \, \mathrm{Z_\odot}$ and $1.5 \, \mathrm{Z_\odot}$ (the 10~Myr bin having a metallicity of $0.2 \, \mathrm{Z_\odot}$ to match the inferred metallicity of the variable SFH bin). Nebular emission is modelled self-consistently with a grid of \textsc{Cloudy} \citep{Ferland2017RMxAA..53..385F} models with the ionisation parameter ($-3 < \log_{10} U < -0.5$) as a free parameter. We included a flexible Charlot \& Fall \cite{Charlot2000ApJ...539..718C} dust attenuation prescription with visual extinction and power-law slope freely varying ($0 < A_V < 7$, $0.4 < n < 1.5$), while fixing the fraction of attenuation from stellar birth clouds to 60\% (the remaining fraction arising in the diffuse ISM; \cite{2019MNRAS.483.2621C}). A first-order correction polynomial \citep{Carnall2019MNRAS.490..417C} is fitted to the spectroscopic data to account for aperture and flux calibration effects. The spectro-photometric fit and the corresponding corner plot are shown in Extended Data Fig.~\ref{Bagpipes_corner_plot}. We find that nearly no wavelength-dependant correction is necessary at the blue end of the spectrum, while at the red end a correction of ~15$\%$ is applied. Crucially, we find a very low SFR (consistent with 0) in the last 10~Myr for \targetid, noting that other tested SFH parametrisations, namely the double power-law SFH described in Carnall et al. 2023 \cite{carnall+2023c} and a single-bin constant SFH with flexible beginning and end of star formation return consistent results. And most crucially agree that the galaxy is quenched. We infer that the oldest stellar population is 40~Myr old, which is equivalent to a formation redshift of z$=$7.6. The galaxy has been quenched for 10~Myr, resulting in a short duration of star formation of 20~Myr from the formation of the galaxy to its quenching. 

\subsubsection{\beagle}\label{sec.beagle}

We use the Bayesian analysis tool \beagle \citep{Chevallard2016MNRAS.462.1415C} to fit to the R100/prism spectrum of \targetid. The \beagle code incorporates a consistent modelling of stellar radiation and its transfer through the interstellar and intergalactic media. We model the SFH as an initial delayed exponential with maximum stellar age, $t_{\mathrm{form}} / \mathrm{yr}$, and location of the peak of star formation as free parameters. In order to disentangle the current star formation rate from the integrated property of total stellar mass, we allow for the most recent episode of star formation to be modelled as a constant with free parameters $\mathrm{SFR} / \mathrm{M_\odot\,yr^{-1}}$ and duration, $t_{\mathrm{quench}} / \mathrm{yr}$ (which can vary between $10^7$ and $10^8$~yr). The nebular emission is characterised by the interstellar metallicity, the ionisation parameter, the mass fraction of interstellar metals locked within dust grains, and crucially \fesc (which can vary between 0 and 1). Dust attenuation follows the two-component prescription of Charlot \& Fall  (2000) \cite{Charlot2000ApJ...539..718C}, where we fit for the total effective \textit{V}-band attenuation optical depth (fixing the ratio of \textit{V}-band ISM attenuation to the \textit{V}-band ISM + birth cloud attenuation to 0.4). We additionally fit for stellar metallicity, stellar mass formed and redshift, totalling 12 free parameters. A list of the free parameters and the adopted priors is presented in Extended Data Table~\ref{Beagle_parameters}.

\begin{table}
    \begin{center}
    \caption{Parameters and associated priors set in \beagle.}\label{Beagle_parameters}
    \begin{tabular}{llll}
        \toprule
        Parameter & Prior & Parameter & Prior \\
        \midrule
       
        $\log(\mathrm{SFR} / \mathrm{M_\odot\,yr^{-1}})$ & $\mathrm{Unif.} \in [-4, 4]$ & Stellar metallicity & $\mathrm{Unif.} \in [-2.2, 0.4]$ \\
        $\log(\mathrm{peak\,of\,SFH} / \mathrm{yr})$ & $\mathrm{Unif.} \in [6, 12]$ & ISM metallicity & $\mathrm{Unif.} \in [-2.2, 0.4]$ \\
        $\log(t_{\mathrm{form}} / \mathrm{yr})$ & $\mathrm{Unif.} \in [6, 13^{\dagger}]$ & Ionisation parameter & $\mathrm{Unif.} \in [-4, -1]$ \\
        $\log(t_{\mathrm{quench}} / \mathrm{yr})$ & $\mathrm{Unif.} \in [7, 8]$ & Dust-to-metal ratio & $\mathrm{Unif.} \in [0.1, 0.5]$ \\
        $\log(\Mstar/\MSun)$ & $\mathrm{Unif.} \in [6, 12]$ & Escape fraction \fesc & $\mathrm{Unif.} \in [0, 1]$ \\
        Redshift & $\mathcal{N}(7.3, 0.1)$ & Total $V$-band att. & $\exp(-\hat{\tau}_\textsc{v})$, $\hat{\tau}_\textsc{v} \in [0, 6]$ \\
        
        \bottomrule
    \end{tabular}
    \end{center}
    $\mathcal{N}(a, b)$ is the Normal distribution with mean $a$ and standard deviation $b$.
    
    $\dagger$ In practice, \beagle will not allow the age of the oldest stars to be greater than the time between $z=20$ and the sampled redshift.
\end{table}

\begin{figure}
  \centering
  \includegraphics[width=1\textwidth]{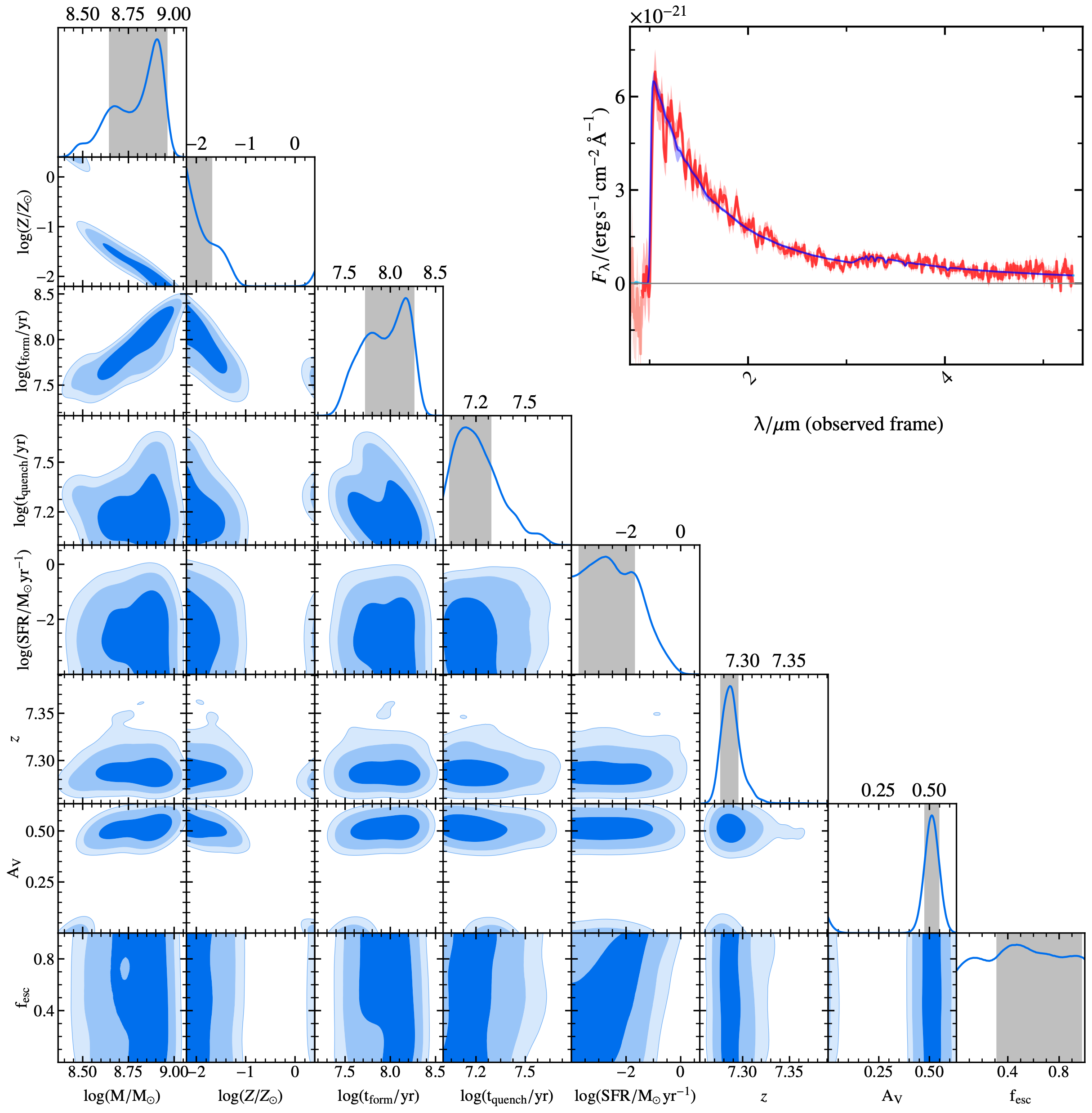}
  \caption{{\bf Summary of key outputs by \beagle.} Bottom left: corner plot with free $\mathrm{f_{esc}}$. Top right: \beagle maximum a-posteriori model of the R100 spectrum.}
  \label{Beagle_corner_plot}
\end{figure}

The corner plot in Extended Data Fig.~\ref{Beagle_corner_plot} shows the \beagle posterior probability distributions of the \beagle fit. The 2D (off-diagonal) and 1D (along the main diagonal) subplots show the posterior distributions on stellar mass \Mstar, metallicity $Z$, SFR, maximum age of stars \tform, minimum age of stars \tquench, redshift $z$, effective dust attenuation optical depth in the V-band $A_V$, and the escape fraction of ionising photons \fesc. The dark, medium and light blue contours show the extents of the 1-, 2-, and 3-\textsigma credible regions.

\beagle gives a current SFR of less than $10^{-1.5}~\MSun\,\mathrm{yr}^{-1}$, a formation time of less than 160~Myr before observation and a quenching time of $\sim$15~Myr before observation. 

We also note that \beagle, as all other three codes, requires some degree of dust attenuation, which suggests some cold gas is still present, which in turn is incompatible with $\mathrm{f_{esc}}\sim 1$. 

\subsubsection{\prospector}\label{sec.prospector}
\begin{figure}
  \centering
  \includegraphics[width=1\textwidth]{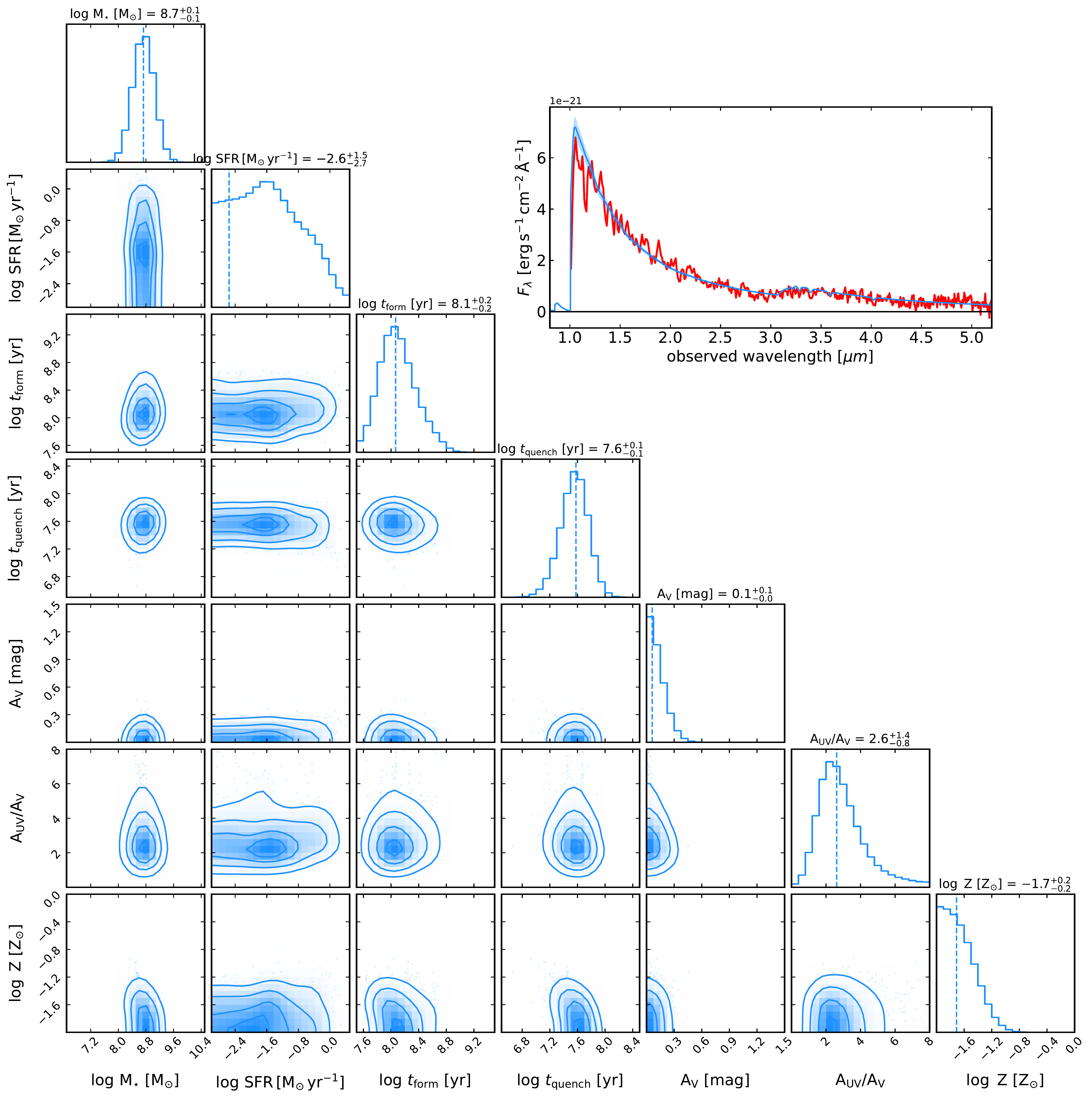}
  \caption{{\bf Summary of key outputs by \prospector.} Bottom left: corner plot with stellar mass \Mstar, SFR, \tform, \tquench, dust attenuation $A_V$, $A_{UV}/A_V$, and stellar metallicity Z. Top right: \prospector maximum a-posteriori model of the R100 spectrum.}
  \label{Prospector_corner_plot}
\end{figure}

We use the Bayesian SED fitting code \prospector \citep{Johnson2021ApJS..254...22J} to model the spectro-photometric data of \targetid. The posterior corner plot for a few key parameters from \prospector is shown in Extended Data Fig.~\ref{Prospector_corner_plot}. The code uses a flexible spectroscopic calibration model, combined with forward modelling of spectra and photometry, to infer physical properties. Following the setup in Tacchella et al. 2022 \citep{Tacchella2022ApJ...926..134T}, we include a flexible SFH (10 bins with the bursty continuity prior), a flexible attenuation law (diffuse dust optical depth with a power-law modifier to shape of the Calzetti et al. (2000) attenuation curve of the diffuse dust \cite{Calzetti2000ApJ...533..682C}), and fit for the stellar metallicity.  Interestingly, \prospector infers a low dust attenuation with $A_V= 0.1^{+0.1}_{-0.0}$ with a rather steep attenuation law ($A_{UV}/A_V = 2.6^{+1.4}_{-0.8}$). This is consistent with the idea that the galaxy has a low gas content and the low SFR in the past 30~Myr before observation. \prospector infers that the oldest stellar population (as defined by the lookback time when the first 10\% of the stellar mass formed) has an age of $\sim$100~Myr, which means a nominal formation redshift of z$=$8.8. The SFR increases significantly $\sim$80~Myr before observation. After this final burst, lasting $\sim$50~Myr, the galaxy quenched on a short timescale. 

We have also experimented with the standard continuity prior \citep{Leja2019ApJ...876....3L}, which weights against sharp transition in the SFH. The overall shape of the SFH is the same, indicating that the data strongly prefers a decreasing SFH in the past $\sim50$~Myr. Quantitatively, the recent SFR (averaged over the past 10 Myr) increases with this prior to $\log_{10}(\mathrm{SFR}/(M_{\odot}\mathrm{yr}^{-1}) = -0.4^{+0.4}_{-0.9}$, still consistent with being quenched and within the uncertainties of the fiducial value obtained with the bursty continuity prior. The quenching time is slightly more recent ($24_{-9}^{+6}$ Myr), but consistent within the uncertainties quoted in Table~\ref{tab:Basic_outputs}.

\subsection{Star-forming, High-\texorpdfstring{\fesc}{f esc} interpretation}

It should be noted that the complete absence of nebular lines always allows, by construction, a solution with $\mathrm{f_{esc}}\sim 1$ (regardless of whether the galaxy has been recently star forming or quiescent) -- the question is whether this solution is accompanied by the production of ionising photons associated with ongoing star formation. 

The fiducial \beagle posterior distribution does not highlight a solution with high \fesc and very recent star formation \citep{Trebitsch+2017}\citep{Zackrisson2017ApJ...836...78Z,Topping2022ApJ...941..153T}. On the contrary, while \fesc is unconstrained, even a value approaching unity indicates a low SFR$< 0.1~\MSun\,{\rm yr}^{-1}$ at 3-\textsigma level (fifth subplot from the left at the bottom of Extended Data Fig.~\ref{Beagle_corner_plot}). 

To assess the very recently star-forming and high \fesc scenario quantitatively, we use \beagle to compare 
two SED models. The model already described (see \ref{sec.beagle}) formally
allows a star-forming solution with high \fesc. The alternative
model has a simplified SFH consisting of a constant SFR; in this
way, low-SFR solutions are effectively removed by the constraint
to form sufficient stellar mass of the appropriate age to reproduce
the observed spectrum.
This alternative model gives $\fesc = 0.98^{+0.01}_{-0.04}$ and $\mathrm{SFR} = 0.63^{+0.05}_{-0.05}~\MSun~\rm{yr}^{-1}$, which is a much higher SFR than the alternative solution.
To select the preferred model, we use the Bayes ratio, i.e. the
ratio between the evidence of the models. The log-difference between the evidences, i.e. the Bayes factor, is $ln(K)=4.1\pm0.3$; according to Jeffreys' criterion \citep{Jeffreys1939}, this is strong evidence for the quenched solution and we adopt it as our fiducial model. 

 As an additional test, we assumed a model with the same setup as the fiducial run, but forcing the escape fraction to $\fesc>0.9$. We find that the result is equal to the fiducial run, and the galaxy remains quenched.

\subsection{Empirical measurements}

To estimate the flux upper limits on \Hbeta and \OIII we sum the
formal variance over three pixels. For $\rm EW_{\Hdelta_A}$ we use
the bands in the Lick definition \citep{worthey+1994}, but without 
any further correction due to spectral resolution. 

We derive an upper limit on the star-formation rate (SFR) from the 3-\textsigma upper limit on the \Hbeta emission-line flux, $F(\Hbeta) \lthan 6.1\times 10^{-20}~{\rm erg} \ {\rm cm}^{-2} \ {\rm s}^{-1}$. To correct this flux for dust
attenuation, we assume the Milky Way attenuation law (Cardelli et al. 1989 \citep{cardelli+1989}), which appears appropriate for galaxies at least up until $z=2.5$ \citep{Reddy2018ApJ...853...56R,Shapley2023arXiv230103241S}. Given that \Hbeta is not detected, we cannot measure the Balmer decrement. We therefore derive the nebular $A_V$ from the continuum $A_V = 0.51$~mag inferred from {\sc Beagle} (the highest value between all models) and upscale this value by 0.64, the median continuum-to-nebular $A_V$ ratio inferred from local
galaxies \cite{Zahid2017ApJ...847...18Z} (of stellar mass comparable to \targetid). The flux is converted to a luminosity assuming
the Planck18 cosmology \citep{Planck2018}.
To convert the \Hbeta attenuation-corrected luminosity to a SFR, we use the conversion factor $2.1\times10^{-42} \, \mathrm{M_\odot \, yr^{-1} \, erg^{-1} \, s}$, appropriate for a Chabrier initial mass function with a high-mass cutoff of 100~\MSun and metallicity $Z=0.27~\mathrm{Z_\odot}$ \citep{Shapley2023arXiv230103241S} (note that this value of
the metallicity is higher than what inferred from the data; this provides a conservative estimate).
This gives a SFR of $0.57 \, \MSun \, \mathrm{yr^{-1}}$. Even stronger constraints come from the \OIII line: we find $F(\OIII)\lthan 6.5\times10^{-20}$~\fluxcgs, which, combined with a conservative assumption \OIII/\Hbeta ratios in high-$z$ galaxies \citep{Cameron2023arXiv230204298C,sanders2023arXiv230106696S}, implies a 3-\textsigma limit on the SFR roughly five times lower than the \Hbeta-derived value (SFR $= 0.12 \, \MSun \, \mathrm{yr^{-1}}$).

Alternatively, assuming the median (and the extreme) observed Balmer decrement 3.5 (5.5) from Shapley et al. (2023) \citep{Shapley2023arXiv230103241S}, we would obtain nebular $A_V$ values of 0.63 and 2.05~mag, respectively. These translate into
\OIII-derived SFRs of 0.10 and $0.34 \, \MSun \, \mathrm{yr^{-1}}$, respectively.
As a comparison, the SFR threshold for quiescence at $z=7.3$ is $0.18 \ \, \MSun \, \mathrm{yr^{-1}}$ (obtained from the threshold
in sSFR defined by $0.2 / t_{\rm H}(z)$ \citep{pacifici+2016} times the {\sc beagle} stellar mass). Thus in all but the
most extreme scenario, \targetid would meet the formal threshold for quiescence.
The absence of emission lines is
independently confirmed by the medium-resolution spectrum (see Extended Data Fig.~\ref{pPXF_R1000}).

\newpage

\noindent{}
\bmhead{Code availability}
The \ppxf, \bagpipes and \prospector codes are publicly available. \beagle is available via a {\sc Docker} image (distributed through {\sc Docker Hub}) upon request at https:/iap.fr/beagle.

\bmhead{Data availability}
The reduced spectra that support the findings of this study are publicly available on GitHub: \url{https://github.com/tobiaslooser/JWST-reveals-a-recently-mini-quenched-galaxy-at-z-7.3}. See MAST at Space Telescope Science Institute for the original data: \url{https://archive.stsci.edu/hlsp/jades}.
\backmatter

\bmhead{Acknowledgements}
We thank the referees for their highly helpful comments, based on which we have substantially improved the discussion and presentation of the results. 
{\small TJL, FDE, RM, JW, WB, LS and JS acknowledge support by the Science and Technology Facilities Council (STFC), by the ERC through Advanced Grant 695671 “QUENCH”, and by the
UKRI Frontier Research grant RISEandFALL. TJL acknowledges support by the STFC Center for Doctoral Training in Data Intensive Science. RM also acknowledges funding from a research professorship from the Royal Society. JW further acknowledges support from the Fondation MERAC. This study made use of the Prospero high performance computing facility at Liverpool John Moores University. BDJ, EE, MR and BER acknowledge support from the \jwst/NIRCam Science Team contract to the
University of Arizona, NAS5-02015. ECL acknowledges support of an STFC Webb Fellowship (ST/W001438/1). SC acknowledges support by European Union’s HE ERC Starting Grant No. 101040227 - WINGS. AJB, JC, AJC, AS, GJC acknowledge funding from the "FirstGalaxies" Advanced Grant from the European Research Council (ERC) under the European Union’s Horizon 2020 research and innovation programme (Grant agreement No. 789056). RS acknowledges support from a STFC Ernest Rutherford Fellowship (ST/S004831/1). SA, BRP and MP acknowledges support from the research project PID2021-127718NB-I00 of the Spanish Ministry of Science and Innovation/State Agency of Research (MICIN/AEI). The Cosmic Dawn Center (DAWN) is funded by the Danish National Research Foundation under grant no.140. H{\"U} gratefully acknowledges support by the Isaac Newton Trust and by the Kavli Foundation through a Newton-Kavli Junior Fellowship. DJE is supported as a Simons Investigator and by \jwst/NIRCam contract to the University of Arizona, NAS5-02015. Funding for this research was provided by the Johns Hopkins University, Institute for Data Intensive Engineering and Science (IDIES). The reserach of CCW is supported by NOIRLab, which is managed by the Association of Universities for Research in Astronomy (AURA) under a cooperative agreement with the National Science Foundation. This research is supported in part by the Australian Research Council Centre of Excellence for All Sky Astrophysics in 3 Dimensions (ASTRO 3D), through project number CE170100013.

We thank Christopher Lovell, Annalisa Pillepich, Deborah Sijacki and Nicolas Laporte for helpful comments and discussions.} 

\bmhead{Author contributions}
TJL, FDE and RM led the writing of the paper. All authors have contributed to the interpretation of the results. TJL, JW, LS, ST, FDE,  ECL, JC, BDJ, WB and KS led the spectro(-photometric) modelling of \targetid and the data visualisation. AB, AD, CNAW, CW, DJE, H-WR, MR, PF, RM, SAl and SAr contributed to the design of the JADES survey. CW contributed to the design of the spectroscopic observations and MSA configurations. KH, ECL, JMH, JL, LW, RE and REH contributed the photometric redshift determination and target selection. AJC, AB, CNAW, ECL, HU, RB and KB, contributed to the selection, prioritisation and visual inspection of the targets. SCa, MC, JW, PF, GG, SAr and BRdP contributed to the NIRSpec data reduction and to the development of the NIRSpec pipeline. SAr, SCh, JC, MC, FDE, AdG, ECL, MM, RM, BRdP, TJL, AS, LS, JS, RS and JW contributed to the development of the JADES tools for the spectroscopic data analysis. BER, ST, BDJ, CNAW, DJE, IS, MR, RE and ZC contributed to the JADES imaging data reduction. CCW, ST, MM, BER, BDJ, CW, DJE, ZJ, JH, AS, KB, AJB, SC, SCh, JC, ECL, AdG, EE, NK, RM, EJN, MJR, LS, IS, RS, KS, HU and KW contributed to the JEMS survey. RHa, BER contributed to the JADES imaging data visualisation. BDJ, ST, AD, DPS, LW, MWT and RE contributed the modelling of galaxy photometry. NB and SAr contributed to the design and optimisation of the MSA configurations. PF, MS, TR, GG, NL, NK and BRdP contributed to the design, construction and commissioning of NIRSpec. MR, CNAW, EE, FS, KH and CCW contributed to the design, construction, and commissioning of NIRCam. BER, CW, DJE, DPS, MR, NL, and RM serve as the JADES Steering Committee.

\bmhead{Author information}
The authors have no competing financial interests. Correspondence and requests for materials should be addressed to Tobias J. Looser (E-mail:
tjl54@cam.ac.uk).

\newpage
\bibliography{bibliography/astrobib}

\end{document}